\documentclass[conference]{IEEEtran}
\IEEEoverridecommandlockouts
\usepackage{cite}
\usepackage{amsmath,amssymb,amsfonts}
\usepackage[linesnumbered, ruled, vlined]{algorithm2e}
\usepackage{algorithmic}
\usepackage{graphicx}
\usepackage{listings}
\usepackage{textcomp}
\usepackage{xcolor}
\usepackage[utf8]{inputenc}
\usepackage[letterpaper, margin=1in]{geometry}
\usepackage{url}

\makeatletter
\@ifundefined{showcaptionsetup}{}{
 \PassOptionsToPackage{caption=false}{subfig}}
\usepackage{subfig}
\makeatother

\usepackage{eso-pic}
\newcommand\AtPageUpperMyright[1]{\AtPageUpperLeft{
 \put(\LenToUnit{0.5\paperwidth},\LenToUnit{-1cm}){
     \parbox{0.5\textwidth}{\raggedleft\fontsize{9}{11}\selectfont #1}}
 }}
\newcommand{\conf}[1]{
\AddToShipoutPictureBG*{
\AtPageUpperMyright{#1}
}
}

\usepackage{tikz}
\usepackage{textcomp}
\usepackage{hyperref}
\newcommand\copyrighttext{%
  \footnotesize \textcopyright 2020 IEEE.  Personal use of this material is permitted.  Permission from IEEE must be obtained for all other uses, in any current or future media, including reprinting/republishing this material for advertising or promotional purposes, creating new collective works, for resale or redistribution to servers or lists, or reuse of any copyrighted component of this work in other works.
  DOI: \href{https://doi.org/10.1109/HPEC43674.2020.9286188}{10.1109/HPEC43674.2020.9286188} }
\newcommand\copyrightnotice{%
\begin{tikzpicture}[remember picture,overlay]
\node[anchor=south,yshift=10pt] at (current page.south) {\fbox{\parbox{\dimexpr\textwidth-\fboxsep-\fboxrule\relax}{\copyrighttext}}};
\end{tikzpicture}%
}

\definecolor{codegreen}{rgb}{0,0.45,0}
\definecolor{codegray}{rgb}{0.5,0.5,0.5}
\definecolor{codepurple}{rgb}{0.58,0,0.82}
\definecolor{backcolour}{rgb}{0.99,0.99,0.97}

\lstdefinestyle{mystyle}{
  backgroundcolor=\color{backcolour},   commentstyle=\color{codegreen},
  keywordstyle=\color{magenta},
  numberstyle=\tiny\color{codegray},
  stringstyle=\color{codepurple},
  basicstyle=\ttfamily\tiny,
  breakatwhitespace=false,         
  breaklines=true,                 
  captionpos=b,                    
  keepspaces=true,                 
  numbers=left,                    
  numbersep=5pt,                  
  showspaces=false,                
  showstringspaces=false,
  showtabs=false,                  
  tabsize=2
}

\SetCommentSty{mycommfont}

\def\BibTeX{{\rm B\kern-.05em{\sc i\kern-.025em b}\kern-.08em
    T\kern-.1667em\lower.7ex\hbox{E}\kern-.125emX}}
\begin{document}
\bstctlcite{IEEEexample:BSTcontrol}

\title{Towards an Objective Metric for the Performance of Exact Triangle Count}

\author{
    \IEEEauthorblockN{Mark P. Blanco\IEEEauthorrefmark{1}, Scott McMillan\IEEEauthorrefmark{2}, Tze Meng Low\IEEEauthorrefmark{1} \\
    \textit{\IEEEauthorrefmark{1}Dept. of Electrical and Computer Engineering ~~~~~ \IEEEauthorrefmark{2}Software Engineering Institute} \\
    \textit{Carnegie Mellon University}\\
    Pittsburgh, PA, United States \\
    \{markb1, scottmc, lowt\}@cmu.edu\\ }
}

\maketitle
\conf{2020 IEEE High Performance Extreme Computing Conference (HPEC)}
\copyrightnotice

\begin{abstract}
The performance of graph algorithms is often measured in terms of the number of traversed edges per second (TEPS). 
However, this performance metric is inadequate for a graph operation such as exact triangle counting. 
In triangle counting, execution times on graphs with a similar number of edges can be distinctly different as demonstrated by results from the past Graph Challenge entries. 
We discuss the need for an objective performance metric for graph operations and the desired characteristics of such a metric such that it more accurately captures the interactions between the amount of work performed and the capabilities of the hardware on which the code is executed. 
Using exact triangle counting as an example, we derive a metric that captures how certain techniques employed in many implementations improve performance. 
We demonstrate that our proposed metric can be used to evaluate and compare multiple approaches for triangle counting, using a SIMD approach as a case study against a scalar baseline.

\end{abstract}


\section{Introduction}
It is widely accepted that software-hardware co-design is required for attaining high performance implementations. 
Therefore, any performance metric used in design and evaluation of an implementation must have properties that bridge the gap between a platform's capabilities and the operations inherent to the problem. 

For many graph algorithms, traversed edges per second (TEPS) is a widely used figure of merit. 
While TEPS suggests that performance is related to the number of edges
that are traversed (read/written during the course of the computation) over time, 
a more common (mis)use of the metric is simply the number of edges in the graph divided by execution time. In many works on triangle counting specifically, the metric used is the simpler definition~\cite{wolf_fast_2017, hu_trix:_2017, samsi_static_2017, mailthody_collaborative_2018, bisson_static_2017}.
From here, we refer to this common usage as edges-per-second.

\begin{figure}
    \centering
    \includegraphics[width=\linewidth]{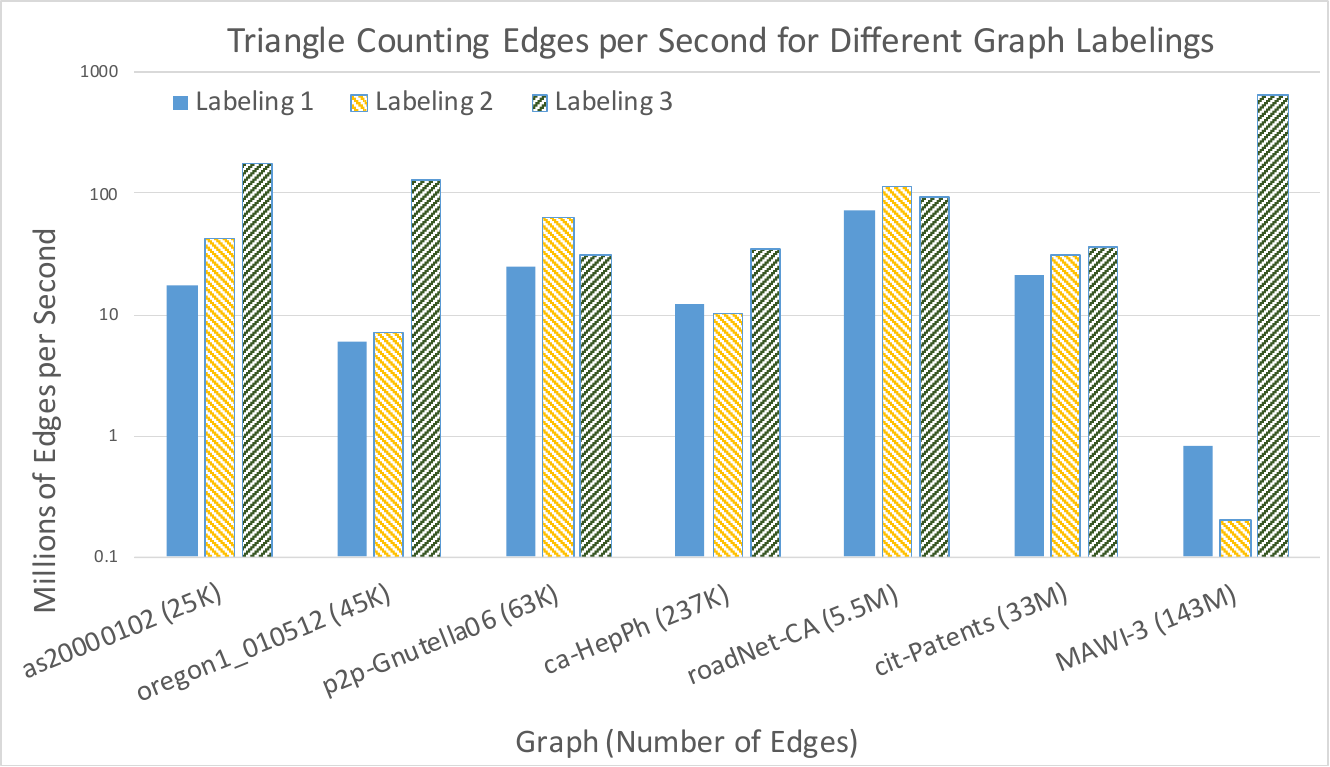}
    \caption{The inadequacy of the edges-per-second performance metric is demonstrated by how the metric varies widely for the each graph that has been sorted in different order, while still retaining the same number of edges, vertices and triangles.}
    \label{fig:motivation}
    \vspace{-7mm}
\end{figure}

To illustrate the inadequacy of the edges-per-second metric for a graph operation such as exact triangle count, consider the plot in Fig.~\ref{fig:motivation} wherein we report the metric performance obtained from a sequential implementation of triangle count for several graphs from the Graph Challenge Dataset~\cite{samsi_static_2017, snapnets}. 
For each graph, we report performance numbers for three different ways in which the vertices have been labeled and sorted.  
Notice that despite having the same number of edges, vertices, and triangles, the performance numbers attained for each graph vary dramatically even when using the edges-per-second metric. 
In essence, the metric is as informative as raw execution time. 
It provides little insight into the software innovation or the hardware capabilities that contribute to the attained performance.


Samsi et. al.~\cite{graphchallenge_analysis1,graphchallenge_analysis2} use edges-per-second and the metric:
\[
T_{tri} = \left(\frac{N_e}{N_1}\right)^\beta,
\]
where $T_{tri}$, $N_e$, and $N_1$ correspond to the execution time, number of edges in a graph, and number of edges processed in one second. In this metric, $\beta$ is a value (smaller is better) representing  technology advancement that arises from implementation, algorithmic, or hardware improvements.
This metric, however, does not explain the difference in execution times seen in Fig.~\ref{fig:motivation}. In addition, it is unclear how this metric can be used to account for the different aspects of technology improvements such as hardware differences and algorithmic improvements.

In this paper, we propose two performance metrics (match checks, and match checks per unit time) that we believe are more objective than edges-per-second. These metrics more accurately describe the performance of triangle counting attained in Fig.~\ref{fig:motivation}, and better capture the relationship between execution time and the expected amount of work that has to be performed. In addition, the proposed metrics expose the relationship between the amount of work and the hardware capabilities required to perform the required work, which in turn provides newer insights into commonly used techniques employed in many triangle counting implementations.

\section{Properties of a Performance Metric}

In this section, we describe properties which we consider desirable in an objective performance metric for hardware-software co-design. We illustrate these properties using dense matrix-matrix multiplication where the use of number of floating point operations per second (FLOPS/s) is a well-established performance metric for dense linear algebra.

\subsection{Indicative of the Expected Amount of Work Performed}
    
    A performance metric needs to measure the expected amount of useful work performed as increasing the amount of work necessarily increases the execution time. By work, we mean the basic unit of computation (and data movement) that is \emph{necessary to compute the desired result using a generally accepted algorithm}. It is necessary to note that the expected amount of work is not the minimum amount of work that could be performed. 
    
    To illustrate this, consider multiplying two matrices of sizes $m \times k $ and $k \times n$. The quantity of work (measured in floating point operations) is approximately $2mnk$ using the traditional triply-nested loop algorithm. Algorithmic innovations such as Strassen~\cite{STRASSEN1969} reduce the expected amount of work, resulting in a faster execution time and thus a higher FLOPS/s when using $2mnk$ floating point operations as the expected amount of work. This metric can potentially report performance above the theoretical peak of the hardware~\cite{jianyuStrassen} because the expected amount of work exceeds that of the actual work.
    
\subsection{Measures Hardware Capabilities}
    The performance metric has to be sufficiently low-level to expose hardware capabilities related to the work being performed. 
    Hardware with more capability for computing the basic work unit should yield a correspondingly increased score using the metric. When normalized to the theoretical capability of the available hardware, the performance metric should also indicate how well the available hardware is being utilized.
    
    
    Increased hardware capabilities such as the introduction of the fused-multiply-accumulate unit (FMA) unit and Single Instruction Multiple Data (SIMD) instruction set extensions can be objectively quantified using FLOPS/s as these hardware capabilities allow more floating point operations to be computed per unit time. Comparing attained FLOPS/s as a function of percentage of theoretical peak hardware capability allows for introspection on implementations to determine if more can be done to achieve better performance.
    
    
\subsection{Captures Implementation Innovations}
\label{subsec:innovations}
    Implementation innovations, such as tiling and data layout changes, increase performance through improved data access while maintaining the amount of work that is performed.  A performance metric should reflect these innovations with a better score despite having no change in both the hardware or the expected amount of work.
    
    A high performance matrix matrix multiplication is often implemented as multiple nested loops that partition a matrix into submatrices through loop tiling/blocking~\cite{goto_anatomy_2008,BLIS1}. In addition, input matrices are repacked in order to ensure that 1) data is brought into the appropriate level of caches and 2) data is repacked such that accesses can be performed with unit stride~\cite{ greghenry}. These implementation choices maintain the same amount of (useful) work, but increase the performance due to better data access. These benefits are demonstrated via a higher FLOPS/s score.
    


 \subsection{Application to Graph Algorithms}
 While we have identified the desirable properties of a performance metric, 
 the vertices of a graph can be relabeled and reordered without changing the structure of the graph. Moreover, the adjacency matrix of the different isomorphic graphs often exhibit different structures that may change the amount of work that needs to be computed. As a convention, we take the original unsorted graph as the canonical graph from which the expected amount of work is computed.  



\section{Deriving a Metric for Triangle Count}
\label{section:metric}

For the sake of completeness, we begin with a brief description of triangle count and commonly used approaches for computing the number of triangles in a graph. We then derive a plausible performance metric for the triangle count operation: match checks.

\subsection{Triangle Counting}
Triangle counting, as its name suggests, counts the number of triangles in a undirected graph $G$. This graph can be represented by its adjacency matrix which is a symmetric sparse matrix.  For the purposes of this paper, we assume that only the lower-triangular portion of the adjacency matrix is stored.

Processing only the lower-triangular (or upper) part of the adjacency matrix allows one to count each triangle exactly once. This approach is common in many implementations~\cite{lee_family_2017, azad_parallel_2015, voegele_parallel_2017}.
Elements in this part of the matrix correspond to edges leading from one vertex to a neighbor with lower vertex ID. This portion of each vertex's neighborhood is referred to as the lower neighborhood, denoted as $N_l(v)$ for vertex $v$.




\subsection {Exact triangle count with wedges and intersections}

\begin{figure}
    \centering
    \includegraphics[width=0.75\linewidth]{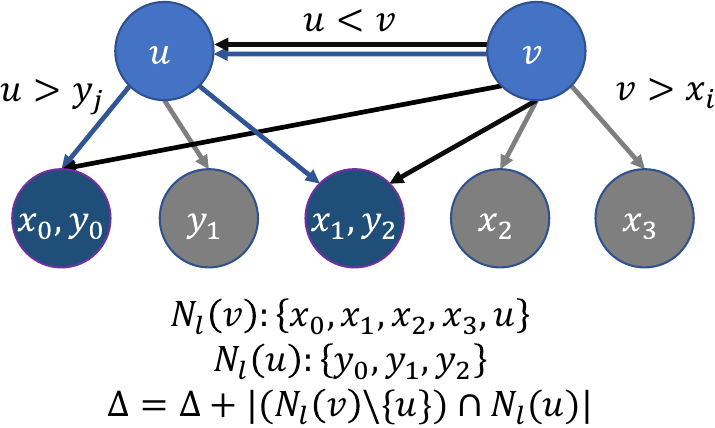}
    \caption{Diagram of connected vertices $v$ and $u$. To count triangles, their lower neighborhoods are intersected, with $u$ subtracted from $N_l(v)$. $\Delta$ is the triangle count.}
    \label{fig:wedge_match_check}
\end{figure}

In general, there are two approaches to counting the number of triangles in a graph. 
The first method is the wedge-check method, where given a wedge (i.e. three vertices connected with two edges), a check is performed to test if there exists an edge that closes the wedge into a triangle. The second approach, based on set-intersection, identifies if there exists a common vertex in the neighborhood of the two vertices of an edge.

We demonstrate the equivalence of the two approaches using Fig.~\ref{fig:wedge_match_check}.
In the following discussion, $v>u$ for both approaches.
In the wedge-check method, the wedges between each pair of vertices $v$, $u$ are found by way of the shared neighbors (e.g. the vertices labeled $x_0/y_0$ and $x_1/y_2$) between their lower neighborhoods ($N_l(v), N_l(u)$). 
The existence of the triangle is then confirmed by checking for the edge directly connecting $v$ to $u$.
In the set-intersection approach, for the endpoints of an existing edge $(v, u)$, their lower neighborhoods are intersected to find shared neighbors (again, $x_0/y_0$ and $x_1/y_2$ in the diagram). Because the intersection is only performed given that edge $(v, u)$ exists, each shared neighbor found in the intersection indicates a triangle.

In practice, work for the first approach is often pre-filtered by existing edges, fusing the two steps and making the two approaches equivalent.
These ways of describing triangle counting mechanics are also compatible with sparse-matrix based approaches. In particular, the work by Wolf et al. for miniTri specifically highlights a wedge-check approach based on a linear algebraic specification~\cite{wolf_task-based_2015}.



From either perspective, the work of finding wedges that may close into triangles constitutes an intersection of $N_l(v)$ and $N_l(u)$.
Algorithm~\ref{alg:scalar_inter} illustrates the core operation in triangle counting by way of a naive approach for set intersection, commonly referred to as merge-based set intersection as in~\cite{inoue_faster_2014, DBLP:conf/icde/ZhangLSF20}.
 
\begin{algorithm}[tb]
\DontPrintSemicolon
\smaller
\KwIn{a1 and a2 initially give the start addresses of each set. a1\_nd and a2\_nd are the addresses just after the end of each set.}
\KwOut{$\Delta$ is the number of matches among both sets.}
\BlankLine
$\Delta \leftarrow 0$ \\
\While{a1 $<$ a1\_nd $\land$ a2 $<$ a2\_nd} {
      \If{*a1 == *a2} {
        a1 $\leftarrow$ a1 + 1\\
        a2 $\leftarrow$ a2 + 1\\
        $\Delta \leftarrow \Delta + 1$\\
      }
      \ElseIf {*a1 $>$ *a2} {
        a2 $\leftarrow$ a2 + 1\\
      }
      \Else {
        a1 $\leftarrow$ a1 + 1\\
      }
  }
  \Return{$\Delta$} \\
  \caption{Scalar merge-based set intersection kernel. Asterisk (*) before an address symbol indicates access to memory.}
  \label{alg:scalar_inter}
\end{algorithm}

\subsection {Expected work for Exact Triangle Count}

Given the previous discussion, the 
core operation in triangle counting can be viewed as finding the size of neighborhood set intersections. Given two neighborhoods of vertices, the first vertex in each of the two neighborhoods are compared (intersected).
In the case of a match, the match is recorded and the next vertices in each neighborhood are compared. Otherwise, the next vertex in the neighborhood of the vertex with the smaller of the IDs is compared with the vertex with the larger ID. This process is repeated until all vertices in one or both neighborhoods have been checked. 

Notice that this work has to be performed even if there are no triangles in the graph since the absence of any triangle can only be ascertained after iterating through the neighborhoods.
This suggests that the number of match checks a graph requires better reflects the amount of work in triangle counting than the number of edges does. 
In addition, the number of match checks performed by triangle counting equals the number of iterations executed by the loop in Algorithm~\ref{alg:scalar_inter}.
From this point in the paper, we refer to such wedge or vertex comparisons as \emph{match checks or matches}.


\subsection{Match Checks in Hardware}
As a metric, match checks do not prescribe which instructions or implementation should be used, but generally indicate that some form of compare instruction followed by conditional updates are required for each match check. Possible implementations of 
Algorithm~\ref{alg:scalar_inter} include 1) a compare operation followed by branches, and 2) predicated instructions to avoid branches~\cite{DBLP:conf/icde/ZhangLSF20}.

However, knowing how match checks are mapped (either manually or by the compiler) to specific instructions or hardware components allows us to use match checks as a proxy for the amount of hardware resources required to compute the necessary amount of work. For example, mapping match checks to a compare instruction naturally limits the rate of match checks to the throughput of the compare (\texttt{CMP}) instruction on a given architecture.




\section{Applying Match Checks as a Metric}
\label{section:sorting_study}
In this section, we demonstrate that match checks are an acceptable metric for work in triangle counting. 

\begin{figure}
    \centering
    \includegraphics[width=\linewidth]{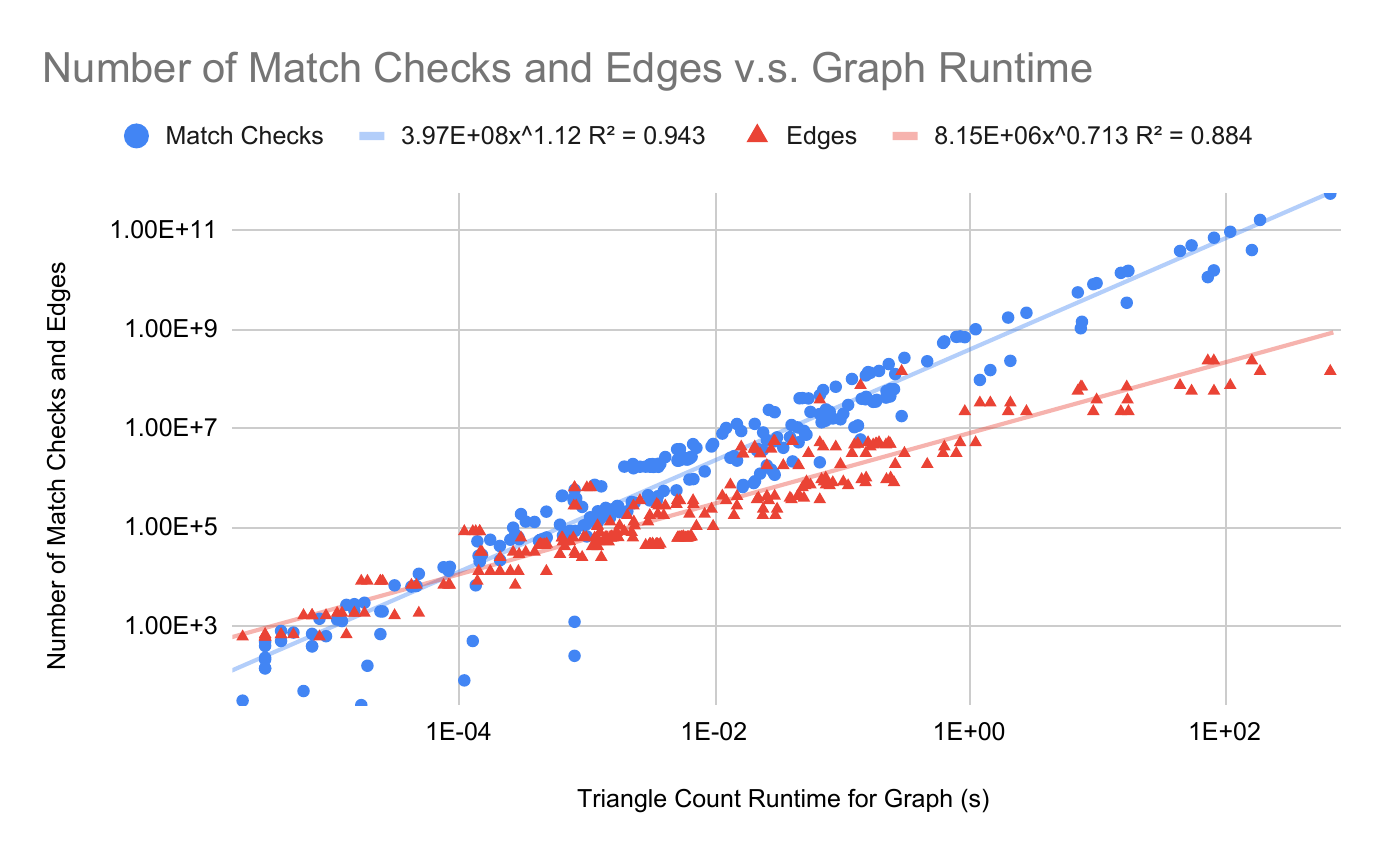}
    \caption{Number of match checks and edges plotted against runtime for scalar merge-sort-based triangle counting, for 3 graph orderings per graph. Increased number of match checks and edges both generally correspond to increased runtime. Match checks show a stronger fit to runtime ($r^2=0.94$) compared to raw number of edges ($r^2=0.884$).}
    \label{fig:runtime_and_checks}
    \vspace{-5mm}
\end{figure}

\subsection{Matches Reflecting Work in Triangle Count}
We validate the use of match checks as a metric of the work for triangle count in Fig.~\ref{fig:runtime_and_checks}.
Each point represents a graph from a real or synthetic graph in the Graph Challenge dataset, processed using a sequential merge-based triangle count implementation~\cite{samsi_static_2017, snapnets}.
The number of match checks per graph was obtained from a second non-timing run, shown in blue.
As expected, the runtime increases with increased number of match checks across the 93 graphs. Hence, match checks are a representative metric for the work in triangle counting. 
By contrast, we also show the number of edges in each graph in red in Fig.~\ref{fig:runtime_and_checks}. While there is clearly some correlation between the number of edges in a graph and the number of checks, it is a weaker one. 


\subsection{Matches Reflecting Implementation Innovation}

It has been observed in numerous works that graph reordering can improve triangle counting performance. 
Chiba and Nishizeki observed that processing vertices in decreasing order of degree enables better running-time bounds for triangle counting based on the graph's aboricity~\cite{chiba_arboricity_1985}.
Recent Graph Challenge implementations such as Bisson et al. attribute improvements in performance to more balanced threads on the GPU~\cite{bisson_update_2018}.
More generally for graph algorithms, Balaji et al. note improvements due to improved memory layout and access patterns~\cite{balaji_when_2018}.
Using the proposed matches-checked metric, we show that new insights can be gained into the need for sorting. 

\begin{figure}
    \centering
    \includegraphics[width=\linewidth]{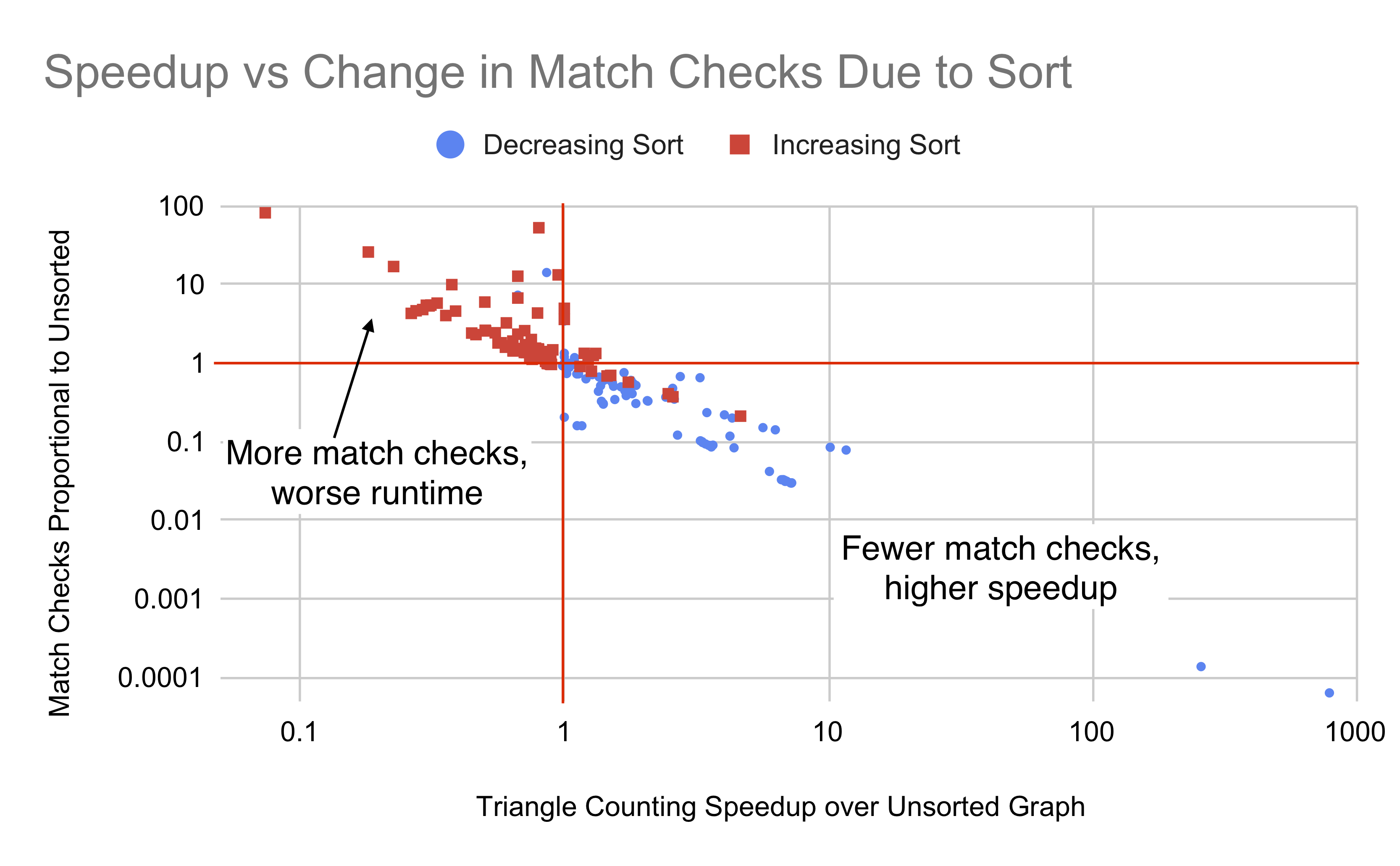}
    \caption{Speedup over non-sorted graph plotted against the change in number of match checks performed for different graph sorts. Different markers represent sort orders. Different sort orders are necessary on different graphs to reduce match checks and execution time.}
    \label{fig:sort_matches}
    \vspace{-5mm}
\end{figure}

\noindent{\em Observation 1: Sorting changes the number of match checks performed.} Besides load-balancing, sorting in the appropriate fashion can significantly reduce the amount of work. 
Fig.~\ref{fig:sort_matches} shows the speedup of triangle counting against the decrease (below 1) or increase (above 1) in proportion of match checks over the unsorted original graph for sequential triangle count. The number of match checks is changed by relabeling vertices based on degree in decreasing order (blue) or increasing order (red). 
For most graphs, we observe that sorting changes the number of match checks performed, and thus a corresponding change in execution time is observed.

{\em Observation 2: Different label orders are required for different graphs.}
While sorting vertices by decreasing vertex degree often leads to lower execution time, a number of graphs benefit from sorting in the reversed order. This is demonstrated in Fig.~\ref{fig:sort_matches} when certain red triangular markers are in the bottom right quadrant, while a small number of blue circular markers are in the top left quadrant.  This suggests that a possible area of future research into heuristics that determine the appropriate sort order or alternative labeling for a given graph.





\subsection{Matches Reflected in Hardware Capability}

\begin{algorithm}[tb]
\DontPrintSemicolon
\smaller
\KwIn{a1 and a2 initially give the start addresses of each set. a1\_nd and a2\_nd are the addresses just after the end of each set.}
\KwOut{$\Delta$ is the number of matches among both sets.}
\BlankLine
$\Delta$ $\leftarrow 0$ \\
\While{a1 $<$ a1\_nd $\land$ a2 $<$ a2\_nd} {
  a1\_max $\leftarrow$ a1[7]\\
  a2\_max $\leftarrow$ a2[7]\\
  \tcp*[l]{Early termination checks}
  a1\_min $\leftarrow$ a1[0]\\
  a2\_min $\leftarrow$ a2[0]\\
  \If{a1\_max $<$ a2\_min}{a1 $\leftarrow$ a1 + 8; continue}
  \If{a2\_max $<$ a1\_min}{a2 $\leftarrow$ a2 + 8; continue}
  vector\_1[0:7] $\leftarrow$ a1[0:7] \\
  \tcp*[l]{packed load}
  vector\_2\_1[0:7] $\leftarrow$ a2[0] \\
  \tcp*[l]{broadcast}
  \tcp*[l]{Omitted: similar broadcasts for a2[1] through a2[7]...}
 
  cmp\_mask\_1 $\leftarrow$ cmp\_eq(vector\_1, vector\_2\_1)\\
  cmp\_mask\_2 $\leftarrow$ cmp\_eq(vector\_1, vector\_2\_2)\\
  cmp\_mask\_3 $\leftarrow$ cmp\_eq(vector\_1, vector\_2\_3)\\
  cmp\_mask\_4 $\leftarrow$ cmp\_eq(vector\_1, vector\_2\_4)\\
  cmp\_mask\_5 $\leftarrow$ cmp\_eq(vector\_1, vector\_2\_5)\\
  cmp\_mask\_6 $\leftarrow$ cmp\_eq(vector\_1, vector\_2\_6)\\
  cmp\_mask\_7 $\leftarrow$ cmp\_eq(vector\_1, vector\_2\_7)\\
  cmp\_mask\_8 $\leftarrow$ cmp\_eq(vector\_1, vector\_2\_8)\\
 
  \tcp*[l]{Omitted: logically or all cmp\_masks together into and\_mask...} 
  $\Delta \leftarrow \Delta$ + popcount(and\_mask) \\ 
  \If{a1\_max $\leq$ a2\_max}{a1 $\leftarrow$ a1 + 8}
  \If{a2\_max $\leq$ a1\_max}{a2 $\leftarrow$ a2 + 8}
  }
  \Return{$delta$}
  \caption{SIMD 8x8 set intersection kernel. Leftover set elements go to the scalar kernel.}
  \label{alg:simd_8x8_inter}
\end{algorithm}

Here we introduce a case study of using match checks to evaluate an alternate approach (SIMD) against a baseline (merge-based scalar) with a focus on hardware-software co-design.
Single-instruction multiple-data (SIMD) hardware is commonly used in regular applications due to the higher throughput it affords.
Use of SIMD in graphs is less common as graph algorithms (inclusive of triangle counting) are considered irregular.
Using match checks, we showcase that SIMD is measurably faster than the scalar approach owing to performing more \textsl{effective} match checks per cycle. 

Recall that match checks can be used as a proxy for the number of times a compare (\texttt{CMP}) instruction is required. The use of SIMD compare instructions (e.g. \texttt{VPCMPEQ}) can increase the number of comparisons that are performed, thus increasing the overall throughput. 

\begin{figure}
    \centering
    \includegraphics[width=0.87\linewidth]{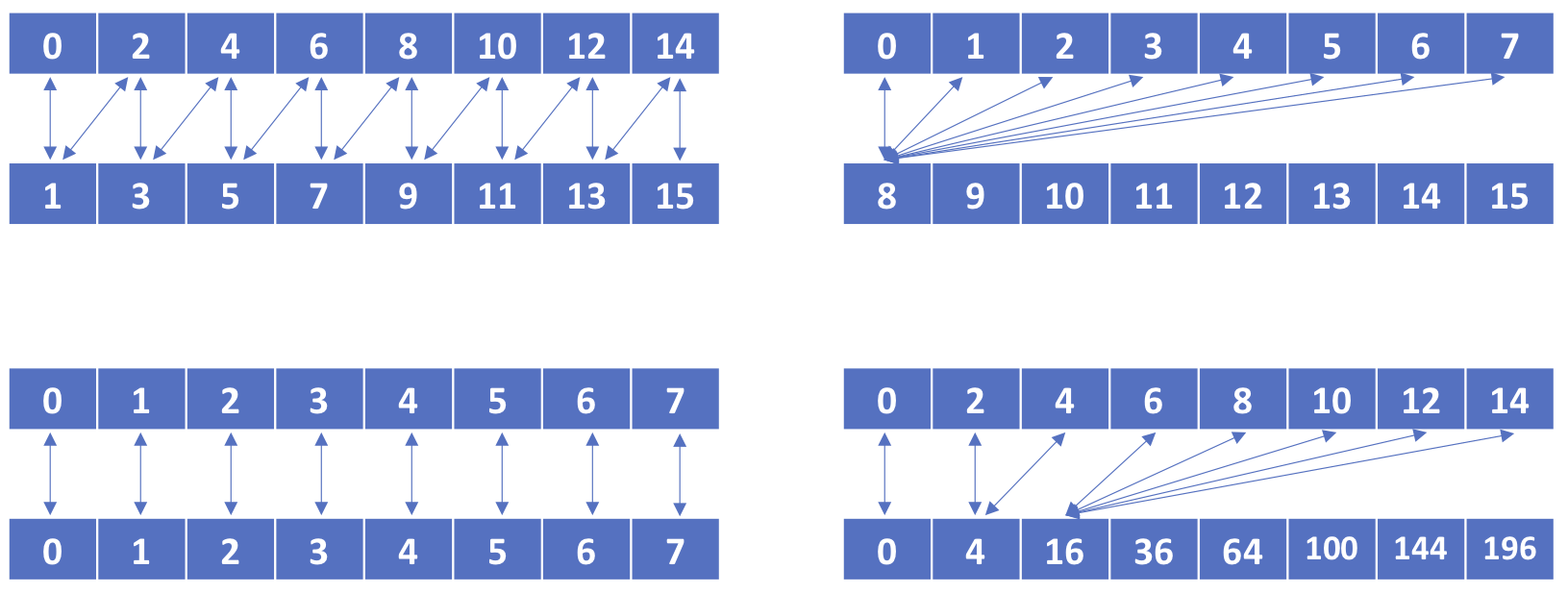}
    \caption{Different distributions of vertex IDs results in different number of match checks. Arrows show the match checks that will be performed by Algorithm 1. The interleaved pattern has the largest number of match checks (15) between two neighborhoods of size 8 each. 
    }
    \label{fig:match_patterns}
    \vspace{-5mm}
\end{figure}

We implemented multiple SIMD set intersection kernels, one of which is shown in Algorithm~\ref{alg:simd_8x8_inter}. This kernel compares eight vertices from each neighborhood against each other. This effectively computes 64 match checks within the kernel. 
However, notice that the number of SIMD match checks corresponding to match checks which the scalar implementation would have performed depends on the distribution of vertex IDs in each vector.
The impact of the different distribution of vertex IDs is illustrated in Fig.~\ref{fig:match_patterns}.
In our analysis, we therefore consider \textit{effective} match checks, where performance of the SIMD approach is measured based on the number of effective match checks it performs relative to the scalar baseline.
The remaining match checks performed in the SIMD approach are considered wasted work.

\begin{figure*}
\begin{tabular}{c}
\includegraphics[width=1\textwidth]{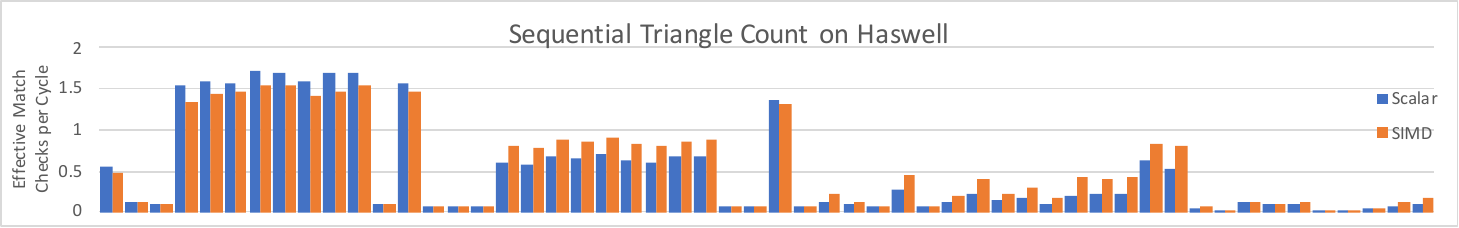}
       \\
\includegraphics[width=1\textwidth]{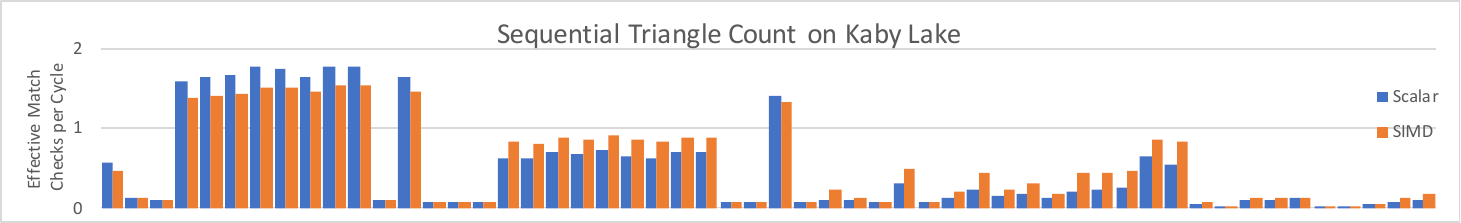}
     \\
\includegraphics[width=1\textwidth]{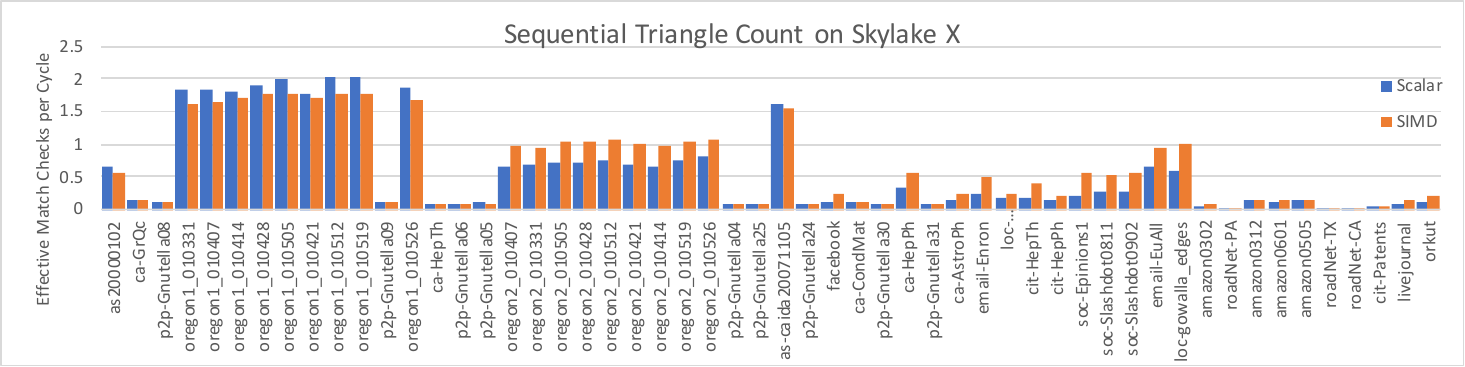}
     \\
\end{tabular}
\caption{Scalar and SIMD triangle count performance, in match checks-per-second, for a variety of graphs from the SNAP dataset on Top) Haswell, Middle) Kabylake, and Bottom) Skylake-X architectures. Use of of SIMD instructions results in higher attained performance for graphs with larger neighborhoods. }
\label{fig:overall}
\end{figure*}

The performance of triangle count using the scalar and SIMD versions of the set-intersection kernels, in match checks-per-cycle, is reported in Fig.~\ref{fig:overall} for the following systems:
\begin{itemize}
    \item[--] Intel Xeon E5-2667 CPU (Haswell)
    \item[--] Intel i7-7700K CPU (Kaby Lake)
    \item[--] Intel Xeon Platinum 8153 CPU (Skylake-X)
\end{itemize}

Sequential performance numbers are shown for all graphs, which are relabeled in decreasing vertex degree order. Sorting time is NOT included in the overall execution time. The graphs in each plots are ordered by increasing graph size (size in memory).

The utility of SIMD set intersection is mixed on the sorted graphs. For smaller graphs, scalar outperforms SIMD. This is likely due to smaller neighborhoods in small graphs reducing opportunities to apply the SIMD kernels. 
The added logic for switching between multiple kernels and the wasted work in the SIMD approach are potential reasons for diminished performance of the SIMD implementation. 
Additionally, the standard deviation of the lower-neighborhood size for most graphs is reduced after sorting, so there are fewer large neighborhoods that SIMD can process.
For larger graphs, in spite of the wasted work, SIMD is often faster than the scalar implementation.
This implementation innovation is reflected in the effective match checks per cycle shown in Fig.~\ref{fig:overall}, particularly for larger graphs.


\label{section:scalar_and_SIMD}

\section{Discussion and Conclusion}
\label{section:conclusion}
In this work, we proposed the use of match checks and match checks per cycle as a performance metric for exact triangle counting. This is motivated by the need for a metric that explains the performance of triangle counting implementations in terms of the amount of work and the hardware capabilities of the system.

We demonstrated that the metric represents the amount of work that has to be performed in triangle counting. The metric also provides new insights into commonly-employed techniques found in many triangle counting implementations such as sort-based relabeling. While we have focused on only sequential implementations, we believe that extending the metric to represent parallel implementations should be straight-forward. 

The observant reader may note that match checks and the number of actually traversed edges in the graph are very similar in numeric value. 
In fact, the actual number of edges traversed equals the number of match checks plus the number of triangles in the graph.
The case we make in this work is not that match checks are better than counting the number of truly traversed edges, but that edges-per-second as used in many recent works is not appropriate and some hardware-focused metric like match checks is needed for hardware-software co-design.

More generally, we believe that the approach we took to identify the proposed metric can be replicated for different (classes of) graph algorithms to identify objective metrics. Having better metrics would provide greater insights into the amount of work to be performed and the hardware capabilities that are required. These insights could lead to faster and more efficient software implementations, and could also suggest hardware features needed by graph algorithms. 

Identified metrics for one graph algorithm could potentially apply to other graph algorithms. For example, the set intersection kernel in triangle counting is similar to a sparse dot product. This suggests that other graph algorithms and sparse linear algebraic workloads may benefit from similar metrics for performance evaluation. We will pursue these directions in future work.

\section*{Acknowledgement}
This material is based upon work funded and supported by the Department of Defense under Contract No. FA8702-15-D-0002 with Carnegie Mellon University for the operation of the Software Engineering Institute, a federally funded research and development center [DM20-0657].

Mark Blanco is supported by the National Science Foundation Graduate Research Fellowship Program under Grant No. DGE 1745016. Any opinions, findings, and conclusions or recommendations expressed in this material are those of the author(s) and do not necessarily reflect the views of the National Science Foundation.

\bibliographystyle{IEEEtran}
\bibliography{references, eps}

\end{document}